\documentclass[11pt,a4paper]{article}

\pdfoutput=1
\usepackage{jheppub}

\usepackage{amssymb}
\usepackage{amsmath}
\usepackage{amsfonts}
\usepackage{graphicx}
\usepackage{color}
\usepackage{xspace}
\usepackage{ulem}
\usepackage{mathtools}
\usepackage{hhline}
\usepackage{float}
\usepackage{caption}
\usepackage{graphicx}

\begin{document}
\title{The inflation trilogy and primordial black holes} 
\date{\today}
\author[a,b]{Paulo B. Ferraz}
\emailAdd{paulo.ferraz@student.uc.pt}

\author[a]{Jo\~{a}o G.~Rosa}
\emailAdd{jgrosa@uc.pt}

\affiliation[a]{Univ Coimbra, Faculdade de Ci\^encias e Tecnologia da Universidade de Coimbra and CFisUC, Rua Larga, 3004-516 Coimbra, Portugal}
\affiliation[b]{Departamento de F\'{\i}sica Te\'orica y del Cosmos, Universidad de Granada, Granada-18071, Spain}



\abstract{
We propose an inflation scenario with three independent stages of cold, warm and thermal inflation, respectively, driven by different scalar fields, motivated by the large number of such fields predicted in most extensions of the Standard Model. We show, in particular, that the intermediate period of warm inflation naturally leads to large density fluctuations on small scales, which can lead to primordial black hole formation in the mass window where they may account for all dark matter. This type of scenario yields a distinctive primordial black hole mass function with a mass gap, with the final period of thermal inflation diluting the abundance of very light black holes.
}

\maketitle


\section{Introduction}
\label{introsection}


Cosmological inflation is one of the pillars of modern cosmology \cite{Guth:1980zm,Linde:1981mu,Albrecht:1982wi,Albrecht:1982mp,Dolgov:1982th,Abbott:1982hn,Linde:1983gd}. The existence of a period of quasi-exponential expansion in the early universe seems to elegantly solve the apparent conceptual problems of the hot Big Bang paradigm, such as the flatness and horizon problems. It also provides the nearly Harrison-Zeldovich spectrum of primordial density fluctuations needed to explain the large-scale structure of the Universe and the observed temperature anisotropies in the Cosmic Microwave Background (CMB) \cite{Planck:2018jri}.
	
Most models of inflation consider that the accelerated expansion is driven by a single scalar field, the inflaton $\phi$ (see e.g.~\cite{Baumann:2009ds} for a recent review). This field follows a classical slow-roll trajectory, with its potential energy dominating over its kinetic energy, i.e. $V(\phi) \gg\dot{\phi}^2/2$, thus leading to a nearly constant Hubble parameter, $H^2\simeq V(\phi)/3M_P^2$, where $M_P$ denotes the reduced Planck mass. Furthermore, (quantum) fluctuations about this homogeneous classical trajectory are stretched and amplified by the quasi-de Sitter expansion, producing an approximately Gaussian and adiabatic distribution of primordial curvature perturbations on super-horizon scales with an almost scale-invariant power spectrum, which provides the seeds for the formation of large-scale structures such as galaxies, clusters, etc. The almost scale-invariant power spectrum has been accurately measured through CMB observations by COBE \cite{COBE:1992syq}, WMAP \cite{WMAP:2003elm} and Planck \cite{Baumann:2009ds}, amongst other ground-based experiments, being characterized by an amplitude $A_s \simeq 2.1\times 10^{-9}$ and a red-tilted scalar spectral index $n_s\simeq 0.965$ \cite{Planck:2018jri}, which means that the correlation between density perturbations at different space points decreases with scale. At the end of inflation, it is conventionally assumed that the inflaton field directly or indirectly decays into the Standard Model degrees of freedom, restoring the hot thermal state assumed in the successful standard cosmological paradigm. This is known as the \textit{reheating} period \cite{Albrecht:1982mp,Kofman:1997yn,Kofman:1994rk}.

This simple \textit{single-field cold inflation} picture is, however, not the only possible way to realize inflation in the early Universe. The plethora of scalar fields predicted in many extensions of the Standard Model, involving e.g.~supersymmetry (SUSY) and/or extra-dimensions, suggest that inflation may follow a non-trivial trajectory in field space, leading to {\it multi-field inflation} scenarios (see e.g.~\cite{GrootNibbelink:2001qt,Wands:2007bd,Dimopoulos:2005ac}). One example  is \textit{hybrid inflation} \cite{Linde:1993cn,Lazarides:2000ck,Bastero-Gil:2006zpr}, where inflation ends with a phase transition triggered by an additional \textit{waterfall field} (as we later describe in more detail).

Another class of inflationary scenarios is known as \textit{warm inflation} \cite{Berera:1995ie,Berera:1995wh, Berera:1996nv,Berera:2008ar}. In these scenarios, the plot for inflation is similar to cold inflation but with an extra actor: a nearly-thermal radiation bath that is sustained by dissipative processes as the inflaton rolls down its potential. These dissipative processes may result from the very same interactions between the inflaton and other degrees of freedom that are responsible for its decay at the end of inflation, and may ease the slow-roll trajectory as long as thermal backreaction effects are sufficiently suppressed. In these scenarios, the radiation bath may, in fact, smoothly take over as the dominant component at the end of the slow-roll period, thus providing a ``graceful exit'' into a radiation-dominated era without the need to include a separate reheating mechanism. Moreover, (classical) thermal field fluctuations typically dominate over their quantum counterparts, leading to a natural enhancement of the primordial curvature power spectrum on large scales \cite{Hall:2003zp, Graham:2009bf,Bastero-Gil:2011rva,Ramos:2013nsa}, while maintaining its approximate scale-invariance in agreement with observations.

Finally, a distinct way of obtaining accelerated expansion is \textit{thermal inflation} \cite{Lyth:1995ka,Lyth:1995hj,Barreiro:1996dx}, typically thought of as a secondary inflationary period after reheating. This scenario is mainly motivated by the large number of scalar flat directions (or flaton fields) predicted in SUSY gauge theories like the MSSM. While the scalar potential is flat along these directions at the renormalizable level in the SUSY limit, it is typically uplifted by non-renormalizable operators and SUSY breaking terms, alongside thermal effects. This naturally leads to cross-over like phase transitions, where a flaton field is stuck in a false vacuum above a critical temperature and rolls down towards a large expectation value as the Universe cools down below the latter. If its potential energy becomes dominant while the field is in the false vacuum, the Universe undergoes an additional period of inflation. This possibility is quite attractive as it may dilute any unwanted cosmological relics, such as moduli fields, gravitinos or topological defects that may be produced during or after reheating.

In the spirit of Occam's razor, single-field cold inflation is typically seen as the simplest way to overcome the shortcomings of standard cosmology. It is, however, not a perfect explanation since a slow-roll evolution is not the natural trajectory followed by a scalar field. In particular, since no symmetries can protect scalar field masses from radiative corrections (apart from axion-like global shift-symmetries, which may, however, be broken by quantum gravity effects), in attempts to realize inflation within fundamental theories like supergravity/string theory one generically finds $\mathcal{O}(H)$ corrections to the inflaton's mass, which prevent slow-roll (see e.g.~\cite{Baumann:2009ds}). This is known as the ``eta-problem''. This problem can be alleviated in warm inflation since the additional dissipative friction may allow for slow-roll with super-Hubble inflaton masses. Strong dissipative effects are, however, typically accompanied by a too large growth of inflaton fluctuations through their interplay with the fluctuations in the radiation bath \cite{Bastero-Gil:2011rva, Bartrum:2013fia, Bastero-Gil:2014jsa, Bastero-Gil:2016qru}. In most cases this leads to a blue-tilted primordial spectrum, or at least a prediction for $n_s$ outside the narrow window favoured by Planck data (see \cite{Bastero-Gil:2019gao} for the only robust exception known). 

One is then led to think that perhaps requiring a single field to drive the full 50-60 e-folds of inflationary expansion may be too much to ask for in a theory of the early Universe. As mentioned above, there is no shortage of scalar fields in most Standard Model extensions. Not of all of them should, of course, be expected to follow slow-roll trajectories or be trapped in false potential minima, given the special conditions that need to be met for these  to occur. But given the three dynamical mechanisms for scalar fields to drive inflation described above, it is not unreasonable to think that all three may have played a role in the early Universe, albeit at different stages. 

Motivated by these considerations, in this work we propose an {\it inflationary trilogy}, i.e.~a scenario with three distinct stages of inflation - cold, warm and thermal - driven by three different scalar fields. The total duration of inflationary expansion is 50-60 e-folds, as required, but each of the three periods accounts for only a fraction of this, thus alleviating the special conditions that need to be met by each scalar field.

The generic picture that we have in mind is as follows. Given generic initial conditions for all the available scalar fields, we assume that there is a direction in field space that is (accidentally or not) free of the eta-problem, i.e.~a scalar field with mass $\lesssim H$ that can drive an initial period of cold inflation. Note that since this period is expected to last less than 50-60 e-folds, the cold inflaton mass can be somewhat larger than in conventional models, thus in principle making it more probable to find such a direction in field space. From the remaining scalar fields, which we assume to have masses $\gtrsim H$ during this period, only those that are (or become) sub-dominant can follow slow-roll trajectories, as we will discuss in more detail.

Once the cold inflaton exits the slow-roll regime, it eventually decays into radiation in the conventional reheating picture. At this stage, the scalar fields that remained in a slow-roll evolution could become dominant, but since they have super-Hubble masses they can only trigger a second period of inflation with the help of dissipative friction effects, i.e.~through interactions with the newly created radiation. As we discussed earlier, dissipation can only sustain slow-roll if thermal backreaction can be avoided. We thus assume that there is a second scalar field, the warm inflaton, endowed with such a mechanism, such as the ones proposed in \cite{ Bastero-Gil:2016qru, Berghaus:2019whh, Bastero-Gil:2019gao, Levy:2020zfo, Ferraz:2023qia}. A second period of warm inflation thus follows, after which either the sustained radiation bath may smoothly take over or a second reheating period brings the universe back into a radiation-dominated regime.

During these first two periods, all other scalar fields are eventually driven to the minimum of the scalar potential. However, for some of these fields thermal backreaction may create false vacua, and their potential energy may come to dominate over radiation for a short period, thus triggering periods of thermal inflation. For simplicity, we assume that only a single period of thermal inflation, driven by a thermal inflaton, occurs, after which the standard radiation era begins, with the usual requirement of a temperature above a few MeV to allow for a successful Big Bang nucleosynthesis (BBN).

We thus have three main actors playing the leading role at different moments in this trilogy, all satisfying the special conditions that allow them to drive inflation in the three possible ways. We could, of course, envisage scenarios with multiple periods of each type, but here we will follow the simplest realization of multi-stage inflation with distinct dynamics in the different stages. 

We will focus on a particular example of this generic picture where, as we will see, a fourth actor plays a supporting role at the start of the trilogy. This is the waterfall scalar field of hybrid inflation mentioned above, since this is one of the few models of cold inflation with less than 50-60 e-folds that is in agreement with Planck measurements of the large-scale primordial curvature power spectrum.

The shape of the primordial power spectrum is, in fact, one of the distinctive features of our proposed inflation trilogy, since the thermal nature of the warm inflaton fluctuations enhances the amplitude of the small-scale curvature perturbations that cross the Hubble horizon during this second inflation period. This enhancement may correspond to several orders of magnitude relative to the large scales that become super-horizon during cold inflation and that we can probe using the CMB, despite the lower potential energy of the warm inflaton field. 

This thus offers a natural scenario for primordial black hole (PBH) formation \cite{Zeldovich:1967lct, Hawking:1974rv, Carr:1974nx}, free of the fine-tuning required to obtain a significant PBH abundance in conventional single-field cold inflation models. As we will show, depending on the relative duration of the three stages of inflation, the resulting PBH mass function may have a peak in the window where these can account for all the dark matter in the present universe (see e.g.~\cite{Escriva:2022duf} for a recent review). Moreover, in our proposed scenario PBHs may form both before and after the final thermal inflation period. In particular, the abundance of the lighter PBHs formed in the period in between warm and thermal inflation is naturally suppressed by the latter, thus predicting a distinctive shape for the PBH mass function.

 This works is organized as follows. In section 2 we present the main features of the inflation trilogy scenario, discussing in detail the dynamics of all the relevant scalar fields in each period, alongside the radiation component, and computing the properties of the primordial curvature power spectrum. We consider a specific realization of the inflation trilogy involving SUSY hybrid inflation \cite{Lazarides:2000ck, Bastero-Gil:2006zpr} for the cold period and the ``Warm Little Inflaton'' scenario \cite{Bastero-Gil:2016qru} for the subsequent warm stage. In section 3 we discuss PBH formation and resulting mass function, determining the parametric regime in which PBHs in the asteroid-mass window accounting for all the dark matter abundance can be produced. We conclude in section 4 with a summary of the main features of our model, discussing its different potential concrete realizations and prospects for future work. Two appendices with technical details are added at the end of this work.

\newpage
\section{Inflationary dynamics}
\label{modelsection}


In this section we discuss in detail the dynamics of the three inflationary stages of our model: cold inflation, warm inflation and finally thermal inflation, in chronological order. As discussed above, these are driven by three different scalar fields that we denote as $\phi_c$, $\phi_w$ and $\phi_t$, respectively. We assume that the fundamental theory includes several other scalar fields but that none of them (except for potentially a waterfall field as we discuss below) plays a significant role in the inflationary dynamics.

\begin{figure}[h!]
\centering\includegraphics[scale=0.35]{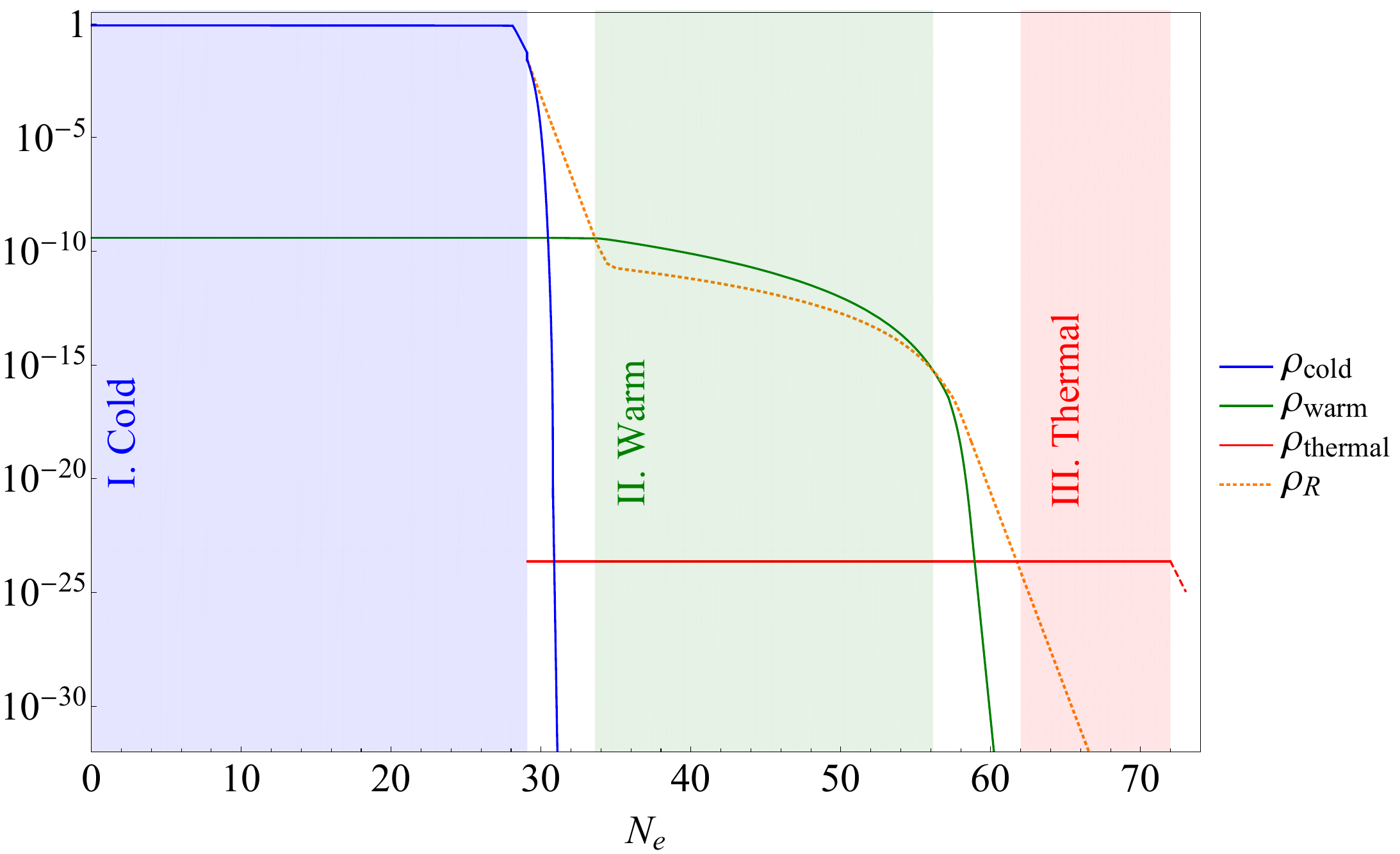}
\caption{Evolution of the energy density of the relevant scalar fields and radiation fluid for our working example of the inflation trilogy setup, highlighting the three stages of cold (blue), warm (green) and thermal (red) inflation (parameter choices given in the text and following figures).}
\label{moneyplot}
\end{figure}

Anticipating our detailed discussion, we show in Figure \ref{moneyplot} an example of the evolution of the energy density of the relevant fields, alongside the radiation energy density that dominates during the periods in between the different inflation stages, as well as the end of inflation when the standard cosmological evolution is attained.

As one can see in this example, the cold inflaton dominates the energy balance at the onset of inflation, after which it decays away into radiation in a first reheating period. During cold inflation, the warm inflaton is slowly rolling down its potential (without thermal friction effects at this stage). The thermal inflaton may be either at the true minimum of its potential or slowly rolling in this first stage. Once a radiation bath is produced, thermal dissipation starts sustaining the warm inflaton's slow-roll trajectory until it eventually comes to dominate and triggers the second, warm inflation stage. Due to the high temperatures attained during reheating and warm inflation, the thermal inflaton is driven to the false minimum of its potential. In this example radiation smoothly takes over the warm inflaton as the dominant component, ending the second inflationary stage. This second radiation era lasts until the thermal inflaton's potential energy becomes dominant. What happens below the critical temperature is somewhat model dependent as we discuss later, potentially including a fast-roll inflation period and/or an early-matter dominated era before the thermal inflaton decays away and radiation takes over, leading to the standard cosmological evolution. Let us now discuss each of these three stages in detail.

\subsection{Cold Inflation}

In this first stage, the field $\phi_c$ dominates, driving a period of cold inflation, while the other fields stay sub-dominant and also in slow-roll (see the discussion below). During this period, the large-scale curvature perturbations that later produce the CMB temperature and polarization anisotropies  are generated and we have the first $N_c$ e-folds of the total $N_{e} \simeq 50-60$ e-folds of inflation required by observations. The resulting power spectrum amplitude $A_s$ and spectral index $n_s$ must match the measured values \cite{Planck:2018jri}
\begin{align}
A_s(k_*) \simeq 2.1\times10^{-9},\quad n_s(k_*) \simeq 0.965,\label{pssi}
\end{align}
at the pivot comoving scale $k_* \simeq 0.05\; \text{Mpc}^{-1}$. The scalar spectral index $n_s$ is generically related to the duration of the inflationary period, thus determining the required number of e-folds of cold inflation. This means that we must consider models predicting $n_s\simeq 0.965$ for $N_c< 50-60$. One of these models is SUSY hybrid inflation \cite{Lazarides:2000ck,Bastero-Gil:2006zpr}, which in its simplest realization considers a superpotential of the form:
\begin{align}
W = \kappa\phi_c(\sigma\overline{\sigma}-M_c^2),
\end{align}
where $\phi_c$ denotes a singlet chiral superfield, while $\sigma$ and $\overline{\sigma}$ are a pair of conjugate superfields and play the role of waterfall fields\footnote{Here we use the same notation for the chiral superfields and their scalar components for simplicity.}. The resulting tree-level scalar potential is given by:
\begin{align}
V_0 = 2\kappa^2|\phi_c|^2|\sigma|^2+\kappa^2(|\sigma|^2-M_c^2)^2.
\end{align}
If initially $\phi_c>M_c$, the waterfall fields are driven to the false minimum at $\sigma =\overline{\sigma}=0$ and the potential is exactly flat. We must, however, take radiative corrections into account, and at 1-loop order these are given by:
\begin{align}
&\Delta V_1 \simeq \frac{(\kappa M_c)^4}{8\pi^2}\mathcal{N}f[x],\\
&f[x] = \frac{1}{4}\Big((x^4+1)\log\frac{x^4-1}{x^4}+2x^2\log\frac{x^2+1}{x^2-1}+2\log\frac{\kappa^2M_c^2x^2}{Q^2}-3\Big),
\end{align}
where $x = |\phi_c|/M_c$, $\mathcal{N}$ is the dimensionality of the representation of the $\sigma$ and $\overline{\sigma}$ fields and $Q$ is the renormalization scale. In this scenario, inflation ends when the cold inflaton field reaches the critical value, i.e.~$x=1$. The waterfall field $\sigma$ then rolls towards the global minimum at $|\sigma|=\sqrt{M_c^2-\phi_c^2}$ and $\phi_c\rightarrow 0$. 

Computing the slow-roll parameters and the duration of inflation in the usual way, it is easy to show that the resulting scalar spectral index satisfies:
\begin{align}
n_s \simeq 1-\frac{1}{N_c}.
\end{align}
which matches the Planck value for $N_c\simeq 28$, thus showing that this is an observationally consistent scenario where cold inflation accounts for only roughly half of the duration of inflation. Note that this is a global SUSY analysis and that supergravity corrections need generically to be taken into account, potentially changing this prediction if one considers the canonical form for the Kahler potential. However, the prediction above can be robust for non-minimal forms of the Kahler potential (see \cite{Bastero-Gil:2006zpr}).

During this period, all sub-dominant scalar fields, and in particular the warm inflaton, follow a slow-roll trajectory if they satisfy the generalized slow-roll conditions (proven in Appendix \ref{appendix1}):
\begin{align}
\epsilon_w<3\Omega_w^{-1},\quad |\eta_w|<3\Omega_w^{-1}
\label{slowrollcold}
\end{align}
where $\Omega_w = \rho_{w}/\rho_{T}$ and $\rho_i$ denotes the energy density in each field $\phi_i$ and $\rho_T\simeq \rho_c$ is the total energy density. Hence, for $\Omega_w\ll 1$ a slow-roll trajectory may be sustained even if the warm inflaton's mass exceeds the Hubble parameter during cold inflation.

\begin{figure}[!h]
\centering\includegraphics[scale =0.57]{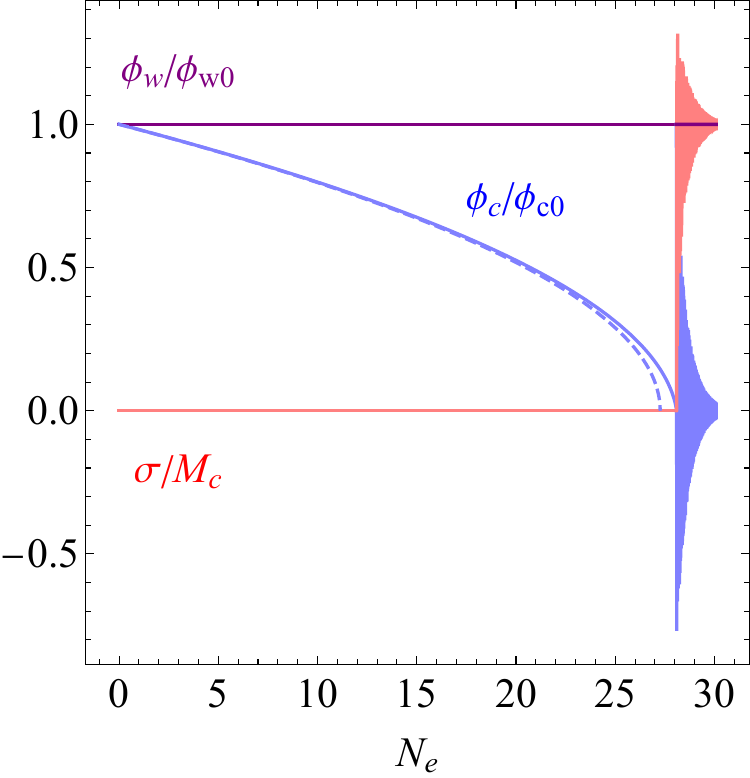}
\caption{Dynamics of the first cold SUSY hybrid inflation stage, in an example lasting $N_c\simeq 28$ e-folds. The solid/dashed curves correspond to the numerical/slow-roll evolution of the cold inflaton (blue), waterfall (red) and warm inflaton (purple) fields, the latter being sub-dominant. We considered $M_c=6.6\times 10^{15}\ \text{GeV},\,\lambda= 7\times10^{-16},\,\kappa= 1,\,\mathcal{N}=1$ and the initial conditions for the fields $\phi_{c,0}=266 M_c,\; \phi_{w,0} \simeq 27 M_c$.}
\label{CI_dynamics}
\end{figure}

To better illustrate this, we show in Figure \ref{CI_dynamics} an example of the evolution of the cold, warm and waterfall scalar fields obtained numerically by solving the corresponding Klein-Gordon equations in a flat FLRW universe alongside the Friedmann equation:
\begin{equation}
\ddot{\phi_i}+3H\dot{\phi}_i+{\partial V\over\partial \phi_i}=0~,
\qquad H^2 = \frac{\sum_i{1\over 2}\dot\phi_i^2+ V(\phi_c,\phi_w,\sigma)}{3M_P^2}
\end{equation}
including a quartic self-interaction term in the scalar potential for the warm inflaton, $\lambda \phi_w^4$.

As one can see in this figure, once the cold inflaton reaches its critical value, inflation ends and the waterfall field rolls down to the global minimum $\sigma=M_c$, with both oscillating around the corresponding potential minima. The warm inflaton remains in a slow-roll evolution, as expected.

In the above solution we have not explicitly included the effects of the decay of the cold inflaton and waterfall fields, which should become significant once they start oscillating about the minimum. In fact, reheating in hybrid models typically involves a tachyonic preheating phase, which efficiently converts a substantial fraction of the energy density of the fields into light degrees of freedom \cite{Garcia-Bellido:1997hex,Garcia-Bellido:1998qth,Bastero-Gil:1999sik}. Due to backreaction of the produced particles, this non-perturbative particle production in general stops before fully converting the cold inflaton and waterfall fields' energy density into radiation, with complete reheating being attained only through perturbative decays. While we will not analyze this reheating phase in detail, since this may be quite model-dependent, in our analysis we convert half of the energy density in the cold inflaton and waterfall fields into radiation within one e-fold of expansion in the oscillating phase, to simulate the effects of preheating (note that the oscillation period is much less than one e-fold) and include a perturbative decay width for both fields to complete the reheating process.


\subsection{Warm Inflation}

Once cold inflation ends and a thermal radiation bath has been produced, the warm inflaton begins exchanging energy with the latter through dissipative effects that may sustain its slow-roll trajectory. Eventually it then becomes dominant over the radiation energy density, thus triggering a warm inflation period. 

During this phase the dynamics is given by the set of equations:
\begin{align}\label{warm_dynamics}
&\ddot{\phi}_w+(3H+\Upsilon)\dot{\phi}_w+V'(\phi_w)=0~,\nonumber\\
&\dot{\rho}_R+4H\rho_R=\Upsilon \dot{\phi}_w^2\\
&H^2 = \frac{\rho_{w}+\rho_R}{3M_P^2}\nonumber~.
\end{align}
Note that, for simplicity, in the above set of equations we have neglected the effects of the cold inflaton and the waterfall fields, assuming that these decay away before the warm inflation phase begins. However, we do include them in our full numerical simulations in the way described in the previous subsection. In these equations, $\Upsilon$ is the dissipation coefficient resulting from the interactions between the warm inflaton and the radiation bath, the latter's energy density being denoted as $\rho_R$. If the radiation is close to thermal equilibrium, its energy density is related to the temperature via $\rho_R=\pi^2g_*T^4/30$ for $g_*$ relativistic degrees of freedom.

Although there may be several ways of realizing warm inflation, here we consider the Warm Little Inflaton (WLI) model \cite{Bastero-Gil:2016qru}. In this scenario, the warm inflaton is the resultant singlet scalar field from the collective spontaneous breaking of a $U(1)$ gauge symmetry. The initial particle content corresponds to two complex scalar fields, $\Phi_1$ and $\Phi_2$ with the same $U(1)$ charge $q$ and an underlying discrete symmetry (see below) ensures that both fields have the same vacuum expectation value $\langle \Phi_1 \rangle = \langle \Phi_2\rangle = M_w/\sqrt{2}$ where $M_w$ is the symmetry breaking scale. The vacuum manifold can be parameterized as
\begin{align}
\Phi_1 = \frac{M_w}{\sqrt{2}}e^{i(\alpha+\phi_w)/M_w},\quad\Phi_2 = \frac{M_w}{\sqrt{2}}e^{i(\alpha-\phi_w)/M_w}.
\end{align}
One can easily check that the overall phase $\alpha$ constitutes the Nambu-Goldstone (NG) boson of the spontaneously broken $U(1)$ gauge symmetry and that in the unitary gauge can be absorbed as the longitudinal component of the now massive gauge field. The relative phase $\phi_w$  remains as a physical scalar degree of freedom after symmetry breaking, being gauge invariant so that we may include in the Lagrangian an arbitrary contribution to the scalar potential $V_w(\phi_w)$. Consistently with the previous sub-section, as an example we choose $V_w(\phi_w)=\lambda\phi_w^4$.

The WLI model also considers interactions between the warm inflaton and other degrees of freedom consistent with the gauge symmetry. Adding two fermions $\psi_1$ and $\psi_2$ whose left-handed components are charged under the $U(1)$ symmetry, while their right-handed counterparts are neutral, we may build the interaction Lagrangian:
\begin{align}
-\mathcal{L}_{\phi\psi} &= \frac{g}{\sqrt{2}}(\Phi_1+\Phi_2)\overline{\psi}_{1L}\psi_{1R}-i\frac{g}{\sqrt{2}}(\Phi_1-\Phi_2)\overline{\psi}_{2L}\psi_{2R}\\
&=gM_w\cos(\phi_w/M_w)\overline{\psi}_{1L}\psi_{1R}+gM_w\sin(\phi_w/M_w)\overline{\psi}_{2L}\psi_{2R},
\end{align}
where we imposed the discrete interchange symmetry $\Phi_1\leftrightarrow i\Phi_2$ and $\psi_1\leftrightarrow \psi_2$. The Dirac masses are therefore oscillatory functions of the inflaton and therefore bounded $|m_{1,2}|\leq gM_w$. The fermions therefore constitute light degrees of freedom provided that during the warm inflation stage the temperature takes values $gM_w\lesssim T \lesssim M_w$. These interactions lead to thermal corrections to the scalar potential, the leading term being of the form $\Delta V_T \propto (m_1^2+m_2^2)T^2$. Since this is independent of the warm inflaton field, it does not contribute to the eta parameter, which is how the WLI model suppresses the troublesome thermal backreaction. However, we have shown in \cite{Ferraz:2023qia} that, even considering a single fermion field and without imposing the discrete interchange symmetry, the thermal backreaction does not prevent warm inflation from occurring, given the oscillatory nature of the corrections to the slow-roll parameters.

Within this setup, the dissipation coefficient $\Upsilon$ can be computed through standard non-equilibrium field theory techniques \cite{Berera:2008ar, Bastero-Gil:2016qru, Levy:2020zfo} and takes the form :
\begin{align}
\Upsilon \simeq C_T T, \quad C_T = \frac{g^2}{h^2}\frac{3}{1-0.34\log h},
\end{align}
where $h$ is the Yukawa coupling between the fermions $\psi_{1,2}$ and its decay products, assumed to be another light fermion $\psi_\chi$ and a light scalar $\chi$. In the concrete realization proposed in \cite{Levy:2020zfo}, $\psi_{1,2}$ are identified with singlet (right-handed) neutrinos (with a Majorana rather than Dirac mass), while $\chi$ is the Standard Model Higgs field and $\psi_\chi$ the known light neutrinos\footnote{We note that realizing this scenario in our present setup is not straightforward, since as we show below our period of warm inflation occurs at a lower scale than in \cite{Levy:2020zfo}, implying lower masses for the would-be right-handed neutrinos. Despite the somewhat lower values of the (neutrino) Higgs Yukawa coupling that we consider here, this may lead to too large Standard Model neutrino masses. Since it is not our goal to fully investigate the parametric regimes yielding an intermediate period of warm inflation, we leave the issue of concrete realizations of our proposal within Standard Model extensions for future work.}. All these fields in the radiation bath may also produce other Standard Model degrees of freedom through thermal scattering processes.

The dynamics after the first reheating then proceeds as follows. Radiation is initially diluted as $\rho_R\propto a^{-4}$ until the dissipative source term $\Upsilon\dot\phi_w^2$ in Eq.~(\ref{warm_dynamics}) becomes comparable to $4H\rho_R$. At this point a slow-roll evolution is attained for both the radiation and the warm inflaton field, with in particular $\rho_R\simeq (3/4)Q\dot{\phi}_w^2$, where $Q=\Upsilon/3H$. Note that radiation is always sub-dominant while the warm inflation slow-roll condition $\epsilon_w<1+Q$ is satisfied, since:
\begin{align}
{\rho_R\over \rho_w}\simeq {\epsilon_\phi\over 2}{Q\over (1+Q)^2}~.
\end{align}
The period of warm inflation ends when this condition is violated, and if $Q\gg 1$ at this stage radiation smoothly takes over as the dominant component, as illustrated in the example of Figure \ref{moneyplot}.

In the WLI setup, the temperature typically falls below the mass of the $\psi_{1,2}$ fermions at the end of the slow-roll regime, and their contribution to the dissipation coefficient becomes Boltzmann-suppressed as $\Upsilon \propto T(m_i/T)^{1/2}e^{-m_i/T}$, where $m_{1,2}$ is the mass of the fermions including thermal corrections \cite{Bastero-Gil:2010dgy,Levy:2020zfo}. While this generically can no longer sustain a slow-roll trajectory, it nevertheless damps the warm inflaton field's evolution as it evolves and then oscillates about the minimum of its potential. 

For a quartic potential, the warm inflaton behaves as a sub-dominant dark radiation component after inflation ends, but eventually its quadratic mass term becomes relevant and it oscillates as cold dark matter. If one assumes the discrete interchange symmetry, this warm inflaton remnant is stable and could even account for all of the dark matter in the present universe \cite{Rosa:2018iff}. Without the discrete symmetry it, however, decays away. While both these possibilities may occur within our 3-stage inflation proposal, we will focus on scenarios where dissipation is quite strong during the warm period, $Q\gg 1$, remaining relatively large after inflation despite the above-mentioned Boltzmann suppression. This means that, whether stable or not, the warm inflaton's contribution to the present-day energy balance is negligible in this regime, but as we will discuss later an alternative explanation for the dark matter puzzle is offered by the intermediate warm inflation period.

\begin{figure}[!h]
\centering\includegraphics[scale= 0.5]{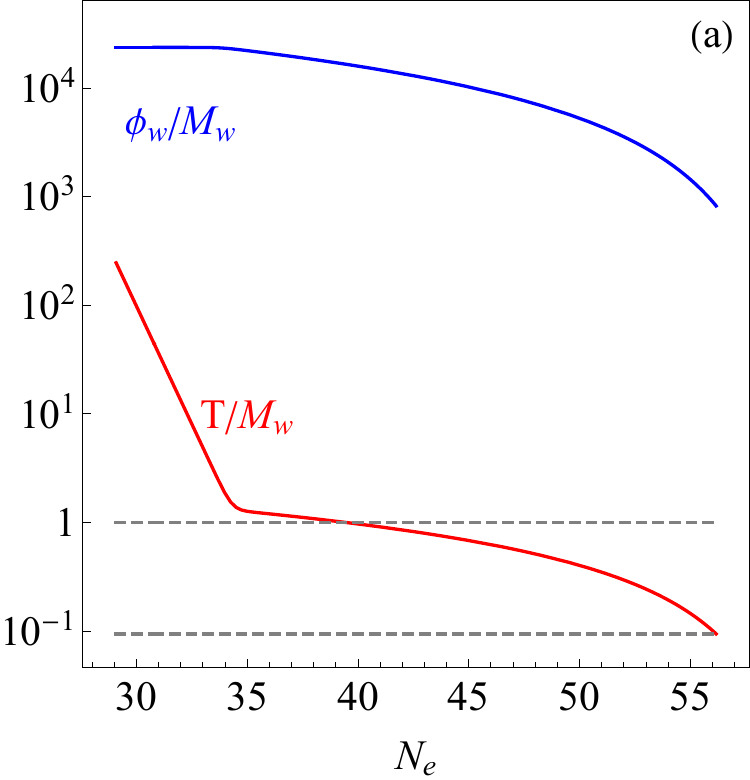}
\centering\includegraphics[scale=0.5]{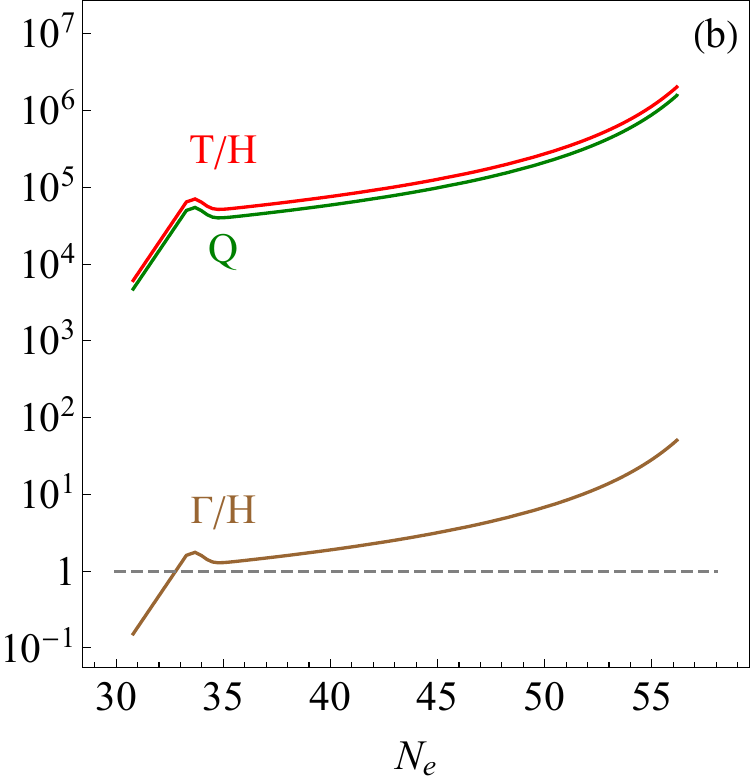}
\includegraphics[scale = 0.5]{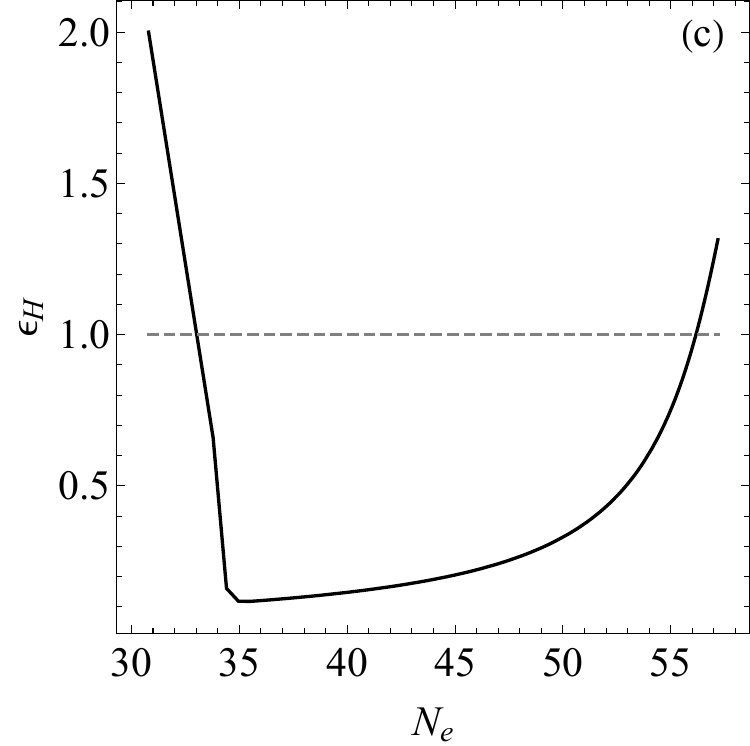}
\caption{Dynamics of the second stage of warm inflation, in an example lasting $N_w\simeq 22$ e-folds after a first radiation era of $N_1=3$ e-folds. We considered $M_w\simeq 5.36\times 10^{12}\ \text{GeV}$, $g\simeq 0.13,\; h \simeq0.21$\; ($C_T\simeq 0.77$)  and $g_*\simeq12.5$ (including $\psi_{1,2},\ \psi_\chi$ and $\chi$). In (a) we show the evolution of the warm inflaton field and radiation temperature; in (b) we show the quantities $T/H$, $Q$ and $\Gamma/H$; and in (c) we show the $\epsilon_H$ parameter, noting that $\epsilon_H<1$ during inflation and $\epsilon_H=2$ during a pure radiation-dominated epoch.}
\label{warmdyn}
\end{figure}

In Figure \ref{warmdyn} we illustrate the dynamics of the warm inflation stage for the same parameters as in Figure \ref{moneyplot}. This figure shows, in particular, the parameter $\epsilon_H = -\dot{H}/H^2$, which takes the value $\epsilon_H\simeq 2$ in a radiation-dominated era and $\epsilon_H<1$ during an inflationary stage. In this example, the warm inflation period lasts for approximately $22$ e-folds, during which a temperature well above the de-Sitter value is sustained by dissipation, $T/H\sim 10^5-10^6$, corresponding to quite strong dissipative effects, since $Q\sim T/H$. This temperature is also below the $U(1)$ symmetry breaking scale $M_w$\footnote{Note that the critical temperature of symmetry breaking  may be $\gtrsim M_w$.} but nevertheless above the (average) fermion masses $m_i\sim g M_w/\sqrt{2}$ for our choice of parameter values.

In this figure we also plot the ratio $\Gamma/H$, where $\Gamma$ denotes the thermally averaged decay width of the fermions $\psi_{1,2}$, computed using the expression given in \cite{Bastero-Gil:2016qru} and considering a Fermi-Dirac distribution \cite{Bastero-Gil:2017yzb}. This shows, in particular, that the fermions remain in a near-equilibrium distribution during the slow-roll regime, as assumed in determining the dissipation coefficient\footnote{The momentum distribution of the fermions produced by dissipation is not exactly thermal, although it  peaks at momenta $p\sim T$. Decays/inverse decays within the thermal bath nevertheless thermalize the fermions if they occur faster than expansion (see e.g.~\cite{Moss:2008lkw}).}.

An important feature of considering an intermediate period of warm inflation is, as anticipated, an enhancement of the amplitude of curvature perturbations for scales crossing the Hubble horizon during this period. This is mainly due to the thermal nature of the warm inflaton fluctuations and their interplay with fluctuations in the radiation bath. The dimensionless curvature power spectrum is thus given by
\cite{Graham:2009bf,Bastero-Gil:2011rva,Ramos:2013nsa}:
\begin{align}
&\Delta_{\mathcal{R}}^2 =\frac{V_w(1+Q)^2}{24\pi^2M_P^4\epsilon_w}\Big(1+2n+\frac{2\sqrt{3}\pi Q}{\sqrt{3+4\pi Q}}\frac{T}{H}\Big)G(Q)
\end{align}
where all quantities are evaluated when the relevant scale crosses the horizon, $k=aH$. Here $n$ denotes the warm inflaton occupation number, which depends on whether the corresponding particles are produced in the thermal bath, although in practice this contribution is negligible for strong dissipation, $Q\gg1$, where the last term within brackets dominates. The function $G(Q)$ encodes the effects of the interplay between warm inflaton and radiation fluctuations and has to be evaluated numerically depending on the shape of the dissipation coefficient and, to some extent, of the scalar potential as well.
For a linear dissipation coefficient $\Upsilon\propto T$ and a quartic scalar potential, this function has been found to be of the form \cite{Bastero-Gil:2016qru}:
\begin{equation}
G(Q) \simeq 1+0.0185Q^{2.315}+0.335Q^{1.364}~.    
\end{equation}
This means that, for $Q\sim T/H\gg 1$, the power spectrum is enhanced by nearly a factor $Q^4$ relative its quantum counterpart. One needs to take into account, however, that the slow-roll parameter $\epsilon_w$ takes larger values during warm inflation, since $\epsilon_H\simeq \epsilon_w/(1+Q)$ in the slow-roll regime when dissipation is taken into account. Also, we must note that the scale of warm inflation is substantially lower than the scale of the first period of cold (hybrid) inflation. Nevertheless, in our example we find $\Delta_R^2\sim 10^{-2}$ on scales crossing the horizon during warm inflation, an increase of nearly seven orders of magnitude relative to the large-scale perturbations generated during the cold inflation period.

While this enhancement is generic but not necessarily so dramatic in all possible parametric regimes realizing an intermediate warm inflation period, we will focus on this type of scenario since it offers a natural setting for producing a significant abundance of primordial black holes that, as we will later show, may potentially explain all of the dark matter.


\subsection{Thermal Inflation}


Upon the second ``reheating'', when radiation smoothly takes over after warm inflation, the universe enters a second radiation-dominated epoch. In our illustrative example, this starts at a temperature  $T_{R2}\sim10^{11}$ GeV, and we may expect in general this to be a temperature well above the electroweak scale. This implies that any remaining light scalar fields should be stabilized by thermal effects near the origin at this stage, even if the true minimum of their potential is non-trivial.

Our thermal inflaton is one of such fields \cite{Lyth:1995ka}, potentially corresponding to a SUSY flat direction \cite{Gherghetta:1995dv, Dine:1995kz} lifted by non-renormalizable terms and SUSY breaking effects (although we note that thermal inflation is not exclusive of SUSY models). The relevant scalar potential near the origin is of the form \cite{Kapusta:2006pm}:
\begin{align}
V(\phi_t, T) \simeq V_0+\frac{1}{2}(\alpha^2T^2-m^2))\phi_t^2+\ldots
\end{align}
where $\alpha$ is an effective coupling depending on how the field interacts with the particles in the thermal radiation bath, and the dots denote higher-order non-renormalizable terms. Note that these thermal mass corrections have a similar origin as those potentially affecting the warm inflaton, with the difference that in the WLI setup these may be cancelled to leading order or are oscillatory in nature.

The zero-temperature field mass term is assumed to be tachyonic, signaling a non-zero vacuum expectation value. The field is then stuck at the origin above a critical temperature $T_{crit}\equiv m/\alpha$, with a constant false vacuum energy $V_0$.

Since the entropy production that characterized the previous warm inflation period is no longer significant, radiation simply redshifts as $a^{-4}$, so that eventually $\rho_R< V_0$. This may occur at a temperature:
\begin{align}
T_t = \left(\frac{30}{\pi^2 g_*}\right)^{1/4}V_0^{1/4}~,
\end{align}
which we assume to be lower than $T_{R2}$, so that the second radiation epoch has a duration of $N_2=\log(T_{R2}/T_t)$ e-folds. The universe thus enters the third inflation stage dominated by the false vacuum energy, $V_0$. This thermal inflation period lasts for a number of e-folds given by:
\begin{align}
N_t\simeq \log \frac{T_t}{T_{crit}} = \log\frac{V_0^{1/4}}{m}+\frac{1}{4}\log\left(\frac{30}{\pi^2g_*}\right)+\log\alpha~,
\end{align}
provided that $T_t> T_{crit}$. We note that accelerated expansion does not necessarily cease below the critical temperature since, even though the field starts rolling towards its true minimum at $\phi_{t0}^2\sim V_0/m^2$, its kinetic energy may remain sub-dominant compared to its potential energy until it is very close to the latter. This period of {\it fast-roll} \cite{Linde:2001ae} may thus increase the effective duration of thermal inflation \cite{Dimopoulos:2019wew}.

Taking this into account and the fact that, for flat directions lifted by non-renormalizable operators, we typically have $\phi_{t0}\gg m$ (or equivalently $V_0^{1/4}\gg m$), it is reasonable to expect a period of $\sim 5-10$ e-folds of thermal inflation to occur after the second reheating, thus helping to dilute any unwanted thermal relics that may have been generated at the end of warm inflation (and that were not, thus, diluted away by the latter). In our subsequent discussion we will treat the duration of thermal inflation as a free parameter, noting that in our working example cold and warm inflation already account for $N_c+N_w\simeq 50$ e-folds of accelerated expansion, so that indeed $N_t\lesssim 10$.

We note that the transition from thermal/fast-roll inflation into the standard radiation-dominated era may be non-trivial, since as $\phi_t$ develops a large expectation value it typically gives large masses to its potential direct decay products. It may nevertheless indirectly decay into light degrees of freedom through virtual-mode exchange, which however suppresses its decay width \cite{Lyth:1995ka}. This means that the thermal inflaton may still dominate the energy balance as cold dark matter for some period until it finally decays away. For simplicity, we will nevertheless consider a radiation-dominated epoch following thermal inflation in our subsequent discussion, since the most important fact is that expansion becomes once more decelerated, while keeping in mind that an early-matter era may occur also within our setup.


\section{Primordial Black Holes}
\label{pbhsection}

We have shown in the previous section that the primordial curvature perturbation power spectrum is greatly enhanced for the scales that cross the horizon during the intermediate warm inflation period, relative in particular to those that become super-horizon during the initial cold inflation stage and generate the large-scale CMB anisotropies. We may thus naturally expect a corresponding increase in the number of large overdensities that may potentially collapse directly into PBHs upon horizon reentry. 

\begin{figure}[!h]
\centering\includegraphics[scale=0.4]{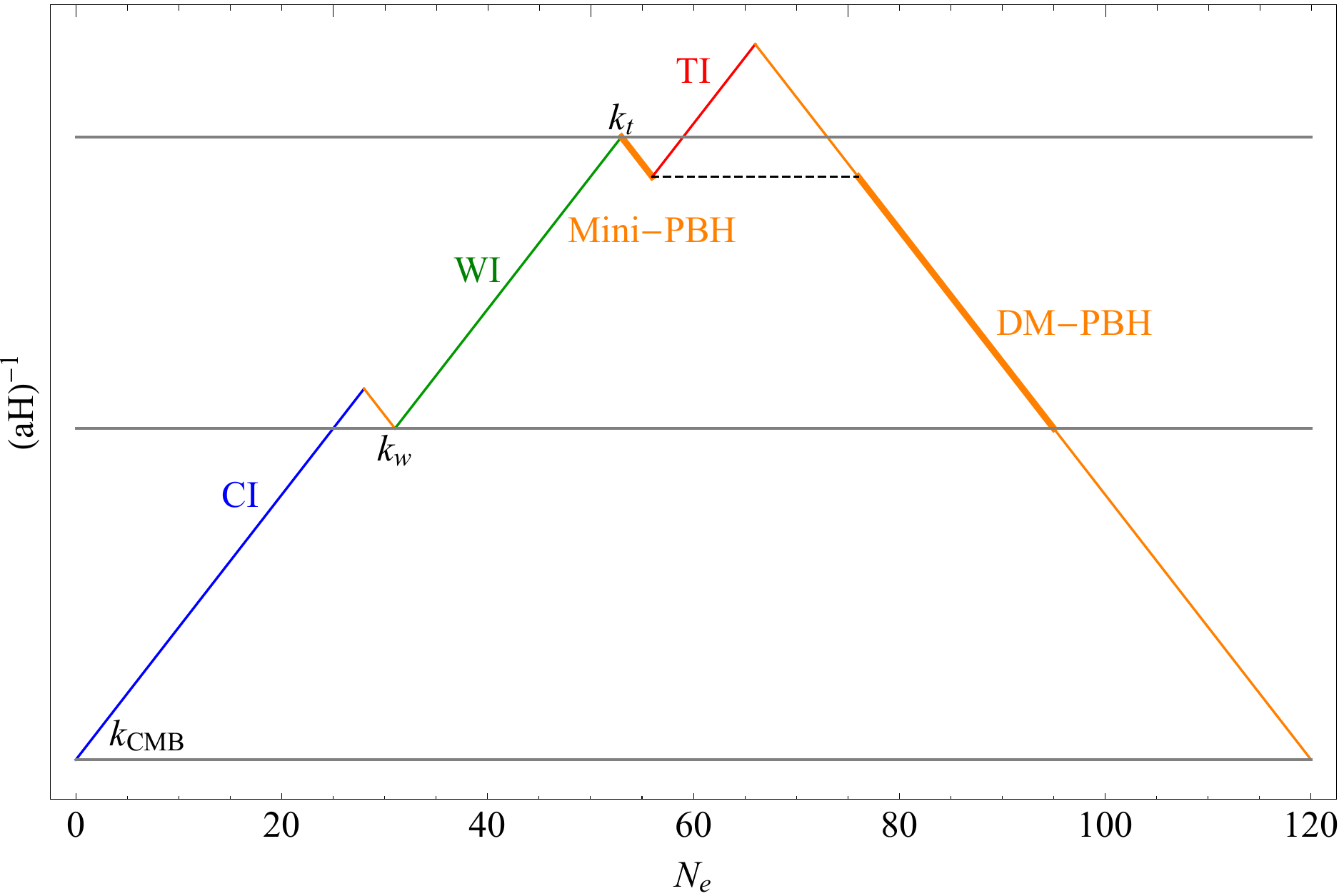}
\caption{Evolution of the Hubble horizon in our working example of the inflationary trilogy scenario. The growing lines correspond to the stages of cold (blue), warm (red) and thermal (green) inflation, while the decreasing lines represent the radiation-dominated eras (dashed orange). The thicker orange lines indicate the periods of PBH formation before (mini-PBH) and after (DM-PBH) thermal inflation. We highlight with the solid horizontal lines the comoving momentum modes $k_\mathrm{CMB}$, $k_w$ and $k_t$, and with the dashed line the comoving mode that separates the scales leading to mini-PBHs and DM-PBHs. The inflationary stages last for $N_c\simeq 28$, $N_w \simeq 22$ and $N_t \simeq 10$ e-folds and the intermediate radiation eras last for $N_1 \simeq N_2 \simeq 3$ e-folds. For illustrative purposes we plot the last 54 e-folds as a single radiation-dominated epoch, although cold dark matter and later the cosmological constant dominate the last $ \simeq 8$ e-folds of expansion before the present day.}
\label{mp2}
\end{figure}

To determine the resulting PBH spectrum, it is crucial to first analyze when different scales re-enter the horizon, since warm inflation is not simply followed by radiation-domination and the additional period of thermal inflation must be taken into account. In Figure \ref{mp2} we show the evolution of the inverse Hubble horizon $aH$ with the number of e-folds. This quantity increases approximately like $a$ during the three inflation stages and decreases as $a^{-1}$ in the three radiation-dominated periods.

In this figure we highlight the comoving scales $k_w$ and $k_t$, such that scales $k_w < k < k_t$ become super-horizon during warm inflation, and thus yield an enhanced power spectrum. Within this interval, we see that, on the one hand, the smallest scales re-enter the horizon in the second radiation era; on the other hand, the largest scales cross-the horizon in the final radiation-dominated era. This then leads to two distinct populations of PBHs.

The mass of a PBH is of the order of the mass within the Hubble horizon when the corresponding overdense region reenters the horizon and collapses \cite{Carr:1975qj, Motohashi:2017kbs}: 
\begin{align}
M_{PBH} = \gamma\frac{4\pi M_P^2}{H},
\end{align}
where $\gamma$ encodes the efficiency of the collapse. For PBHs formed in a radiation-dominated era, $\gamma\simeq 0.2$ \cite{Carr:1975qj, Inomata:2017okj, Sasaki:2018dmp, Mishra:2019pzq}, and we then have:
\begin{equation}
M_{PBH} = 1.55\times 10^{24}M_{\odot}\Big(\frac{\gamma}{0.2}\Big)\Big(\frac{g_*}{106.75}\Big)^{1/6}(1+z)^{-2}\xi^2,
\label{bhmass}
\end{equation}
where $M_\odot$ is the mass of the Sun, so that lighter PBHs form earlier. The factor $\xi=1$ for PBHs formed in the final radiation era, while $\xi=e^{N_t}$ for those born before the onset of thermal inflation. This is due to the fact that in both radiation eras we have $H\propto a^{-2}=(1+z)^2$ but taking into account that the scale factor increases by $e^{N_t}$ after thermal inflation. This can also be seen by noting that the last comoving scale to reenter the horizon before thermal inflation is nearly the same as the first to come into the horizon in the final radiation era (denoted by the dashed horizontal line in Figure \ref{mp2}), with $H$ decreasing by $e^{-2N_t}$ in between these two horizon-crossing moments.

We therefore obtain a mass gap of $\xi^2=e^{2N_t}$ between the two populations of PBHs that may form in our setup. On the one hand, in our working example, the {\it mini-PBHs} formed before thermal inflation have masses in the range $7\times 10^{4}\ \text{kg}\lesssim M_{PBH}<3\times 10^{7}$ kg. Since these are lighter than $5\times 10^{11}$ kg, they evaporate in less than the age of the Universe and, hence, do not contribute to the present dark matter abundance. On the other hand, PBHs formed after thermal inflation are much heavier, with masses in the range $2\times 10^{16}\ \text{kg}<M_{PBH}<6\times 10^{32}\ \text{kg}$. Given that after formation their energy density simply redshifts as cold dark matter, we denote them as {\it DM-PBHs}\footnote{Note that here we are assuming that the curvature perturbations generated during thermal inflation are not large enough to generate a significant PBH abundance. While the analysis in \cite{Dimopoulos:2019wew} suggested that thermal inflation could lead to efficient PBH production, it did not take into account the effects of thermalization of field fluctuations, which as shown in \cite{Bastero-Gil:2023sub} may significantly suppress the curvature power spectrum.}.

The present fraction of dark matter in the form of PBHs, $f_{PBH}=\rho_{PBH,0}/\rho_{DM,0}$ is related to their fractional energy density at formation, $\beta_{PBH}=\rho_{PBH}/\rho_{tot}$, via (see e.g.~\cite{Sasaki:2018dmp}):
\begin{align}
f_{PBH}=\Omega_{DM,0}^{-1}\left(\frac{H}{H_0}\right)^2(1+z)^{-3}\beta_{PBH}~.
\label{bhmassfraction}
\end{align}
Using Eqs.~(\ref{bhmass}) and (\ref{bhmassfraction}), we obtain:
\begin{align}
f_{PBH} =1.68\times 10^8\left(\frac{\gamma}{0.2}\right)^{1/2}\Big(\frac{g_*}{106.75}\Big)^{-1/4}\left(\frac{M_{PBH}}{M_\odot}\right)^{-1/2}\beta_{PBH}\xi^{-3}.
\end{align}
so that the abundance of mini-PBHs is suppressed by a factor $e^{-3N_t}$, which reflects the dilution effect of thermal inflation. Note that, for the mini-PBHs, $f_{PBH}$ denotes the dark matter fraction one would obtain today had the PBHs not evaporated. As we will see below, the fact that mini-PBHs are diluted away by thermal inflation is quite important for the success of this setup, since PBHs with $M>10^6$ kg evaporate after the epoch of Big Bang Nucleosynthesis (BBN), and even a small abundance could spoil the latter's predictions.

The fractional energy density at formation $\beta_{PBH}$ can be computed using the Press-Schechter formalism \cite{Press:1973iz} such that, for a given PBH mass, it corresponds to the probability of the Gaussian comoving density contrast perturbation\footnote{Here we neglect, to a first approximation, the effects of non-Gaussianity in the primordial spectrum, although these may play a significant role \cite{DeLuca:2019qsy}.}, coarse-grained over a reference scale (typically taken to be the comoving Hubble horizon), being larger than the threshold $\delta_c=0.4-0.5$ (see e.g.~\cite{Musco:2012au, Harada:2013epa, Motohashi:2017kbs}) for black hole formation\footnote{For concreteness here we consider $\delta_c=0.414$.}:
\begin{align}
\beta_{PBH}= \gamma\int_{\delta_c}^1 P(\delta)d\delta \simeq \gamma\frac{\sigma}{\sqrt{2\pi}\delta_c}e^{-{\delta_c^2\over 2\sigma^2}}~.
\end{align}
The variance $\sigma=\sigma(M_{PBH})$ is obtained from the density perturbation power spectrum via:
\begin{align}
\sigma^2 = \int d\log k \Delta_\delta^2(k)W^2(k,k_{PBH})~,
\end{align}
where $W(k,k_{PBH})$ is the window function used to smear the initial density contrast at the relevant scale to obtain the coarse-grained one. The power spectrum of the density contrast is related to the power spectrum of the comoving curvature perturbation as \cite{Green:2004wb}
\begin{align}
\Delta^2_\delta = \frac{4(1+w)^2}{(5+3w)^2}\Big(\frac{k}{aH}\Big)^4\Delta^2_{\mathcal{R}},
\end{align}
where $w$ is the equation of state parameter for the dominant fluid at that time, e.g. $w=1/3$ for a radiation fluid. We notice that since different perturbation modes re-enter the horizon at different times, we can relate the mass of a black hole to the wavelength mode of the perturbation that originated it, i.e. $k_{PBH} = k_{PBH}(M_{PBH})$.

As mentioned previously, during the intermediate warm inflation stage, the comoving curvature perturbations power spectrum may reach values $\Delta_\mathcal{R}\sim 10^{-2}$. Writing the latter as $\Delta^2_\mathcal{R} =A_s(k/k_w)^{n_s-1}$ and using a Gaussian window function $W = \exp(-k^2/2k_{PBH}^2)$, we obtain upon integration over $k_w<k<k_t$, i.e.~the modes that exit the horizon during warm inflation:
\begin{align}
\sigma^2=\frac{8}{81}A_s\left(\frac{k_{PBH}}{k_w}\right)^{n_s-1}\left(\Gamma\left[\frac{n_s+3}{2},\left(\frac{k_w}{k_{PBH}}\right)^2\right]-\Gamma\left[\frac{n_s+3}{2},\left(\frac{k_t}{k_{PBH}}\right)^2\right]\right)
\end{align}
where $\Gamma[a,z]$ denotes the incomplete gamma function. The spectral index for the warm inflation stage is computed in detail in Appendix \ref{sisection}, yielding in general a nearly scale-invariant and slightly blue-tilted spectrum for strong dissipation, $Q\gg 1$. 

\begin{figure}[!h]
\centering\includegraphics[scale = 0.4]{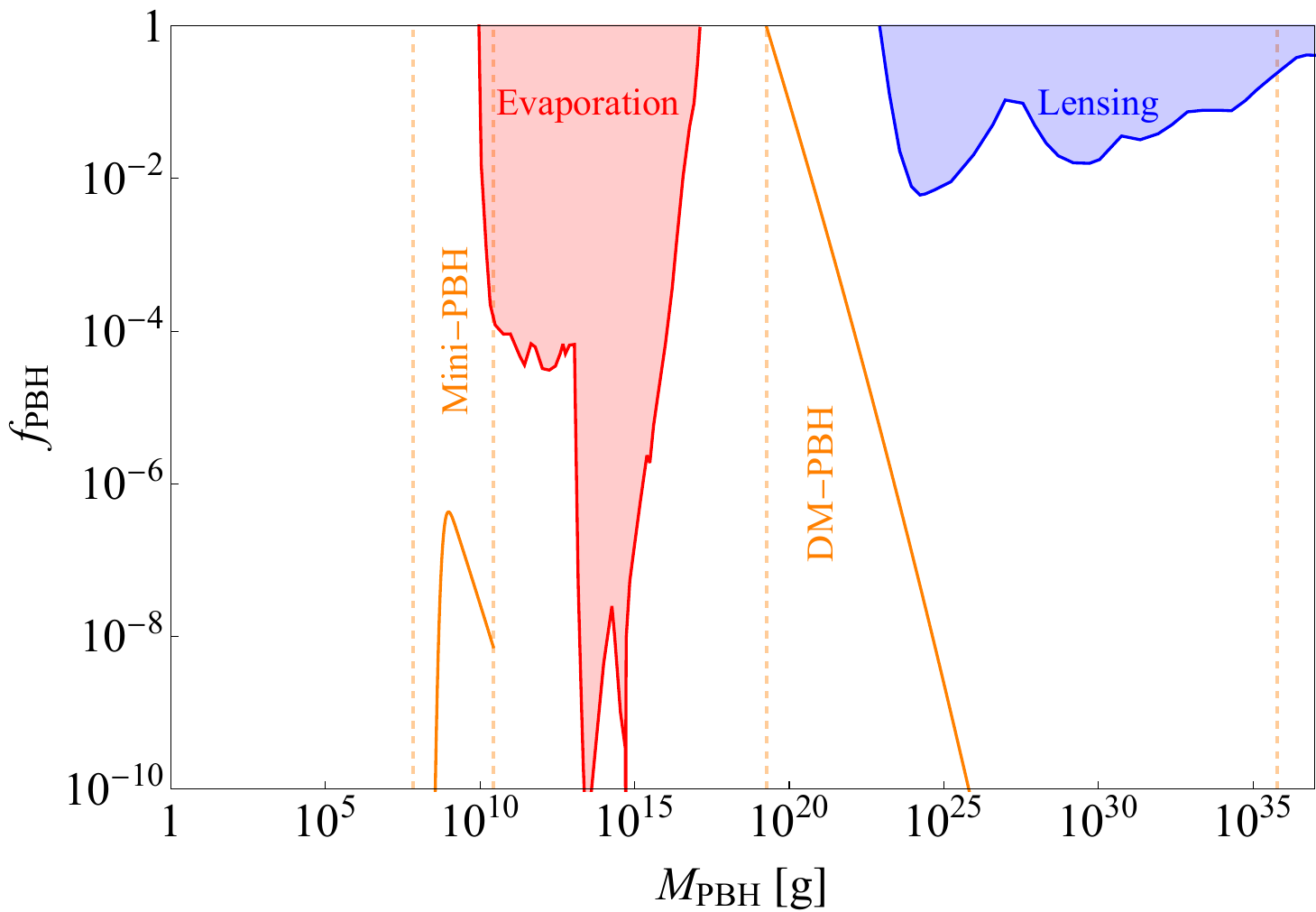}
\caption{Predictions of the inflation trilogy scenario, in our working example, for the fraction of dark matter in the form of PBHs (solid orange curves). The vertical orange lines indicate the limits of the ``mini-PBHs'' and ``DM-PBHs'' mass ranges. The shaded regions give the current constraints on the PBH abundance from Hawking evaporation \cite{Carr:2009jm, Acharya:2020jbv, Chluba:2020oip, Carr:2016hva, Boudaud:2018hqb} (red) and gravitational microlensing \cite{Smyth:2019whb, Niikura:2019kqi, Zumalacarregui:2017qqd} (blue).}
\label{PBH_mass_fraction}
\end{figure}

Using these results, we have computed the resulting PBH mass fraction for our working example of the inflationary trilogy proposal, with 28 e-folds of cold inflation followed by 22 e-folds of warm inflation and then by 10 e-folds of thermal inflation. In this example, the intermediate radiation eras are relatively short, with $N_1=N_2=3$ e-folds. Results are shown in Figure \ref{PBH_mass_fraction}, where we also plot the main constraints from Hawking emission and gravitational microlensing on $f_{PBH}$. 

As one can see in this figure, the lightest PBHs in the DM-PBH mass range (i.e. those formed in the standard radiation era) are dominant, and in this example fall naturally within the mass window where they may account for all the dark matter. The mini-PBHs formed before thermal inflation have a suppressed abundance, well below the bounds imposed by BBN, and in fact most are light enough to evaporate before nucleosynthesis.

One should note that we cannot accurately predict the PBH mass function very near the limits of the mass gap, which correspond to the same comoving mode. This mode is the last to reenter the horizon before thermal inflation and the first to do it in the standard radiation era (depicted with a dashed line in Figure \ref{mp2}). Only a detailed analysis of the dynamical evolution of overdensities at this scale can reveal whether they lead to the heaviest mini-PBHs or the lightest DM-PBHs, and this is left for future work. It is nevertheless clear that the DM-PBH abundance should peak near the lower end of this mass window, so that dark matter could indeed be fully accounted for by asteroid-mass PBHs within in our proposal.

We must also point out that the relative duration of the last two inflationary stages and radiation epochs has a significant impact on the PBH spectrum. A longer warm inflation stage broadens the mass window for both mini-PBHs and DM-PBHs, while a longer period of thermal inflation not only broadens the mass gap but further suppresses the mini-PBH abundance. The duration $N_2 $ of the intermediate radiation epoch has also an important effect on the PBH mass function, with a longer era increasing both the lower and upper limits of the mass gap, i.e.~leading to heavier mini-PBHs and DM-PBHs. We find that, for the duration of the inflationary stages considered in our working example, PBHs in the asteroid-mass range are formed for $N_2\lesssim 7$, but that current constraints on evaporating PBHs (in the mini-PBH window) require $N_2\lesssim 6$, implying a thermal inflation stage starting at a relatively high-temperature ($T_t\gtrsim 10^9$ GeV) and ending above the electroweak scale, which may be relevant for baryogenesis. These bounds would, of course, change if the relative duration of the warm and thermal inflation stages are distinct from our working example, a detailed analysis that is, however, outside the scope of this work.

\section{Discussion and Conclusion}


Inflation is the most widely accepted solution to the shortcomings of the standard cosmological paradigm. While most models consider only a single period of accelerated expansion in the early Universe, driven by the dynamics of a single scalar field, the plethora of scalar fields available in most extensions of the Standard Model motivated us to consider a multi-stage inflation model driven by different fields. Moreover, the three known dynamical mechanisms allowing a scalar field to mimic a cosmological constant suggest considering a three-stage scenario involving cold, warm and thermal inflation.

We envisage a class of scenarios with special directions in scalar field space endowed with mass protection mechanisms. One must have a mass below the Hubble parameter to drive a period of cold inflation on its own; another may have a super-Hubble mass but it must be protected against thermal corrections, in order to drive a period of warm inflation; finally the thermal inflaton's mass need not be protected against thermal effects, but typically it corresponds to a flat direction at the renormalizable level. 

The cold-warm-thermal chronological order of our trilogy scenario follows naturally from the dynamics in scalar field space. Since only the cold inflaton can sustain slow-roll on its own, it naturally comes to dominate over the other scalar fields with $m\gtrsim H$, which evolve towards the minimum of their potential during the first period of cold inflation. However, as we have discussed, once their abundance becomes sufficiently diluted, $\Omega\lesssim \mathcal{O}(10^{-2})$ these scalars may also enter a slow-roll regime. Once cold inflation ends and the universe is reheated for the first time, the slow-roll trajectory of the warm inflation can be sustained by dissipative effects, and once it becomes dominant a new period of warm inflation naturally occurs. If dissipative effects are sufficiently strong, as we have shown to be possible in our illustrative example, radiation smoothly takes over as the dominant fluid at the end of the warm inflation period. At this stage all other scalar fields should have evolved to the minimum of their potential, particularly since they are expected to acquire large thermal masses. For many fields this does not coincide with the zero-temperature minimum, but they nevertheless remain there until the temperature falls below a critical value, thus sustaining a temporary cosmological constant that may come to dominate over radiation, hence driving a third period of thermal inflation.

These arguments may, of course, apply to more scalar fields with these properties, and in principle we could envisage scenarios with more than three stages, although for simplicity we chose the minimal setup realizing the above-mentioned three dynamical realizations. 

As a ``proof-of-concept'', we illustrated our proposal with a concrete example where cold inflation occurs within a supersymmetric hybrid model, where the predictions for the large-scale curvature power spectrum agree with observational Planck data despite the relative short period of cold inflation considered of less than 30 e-folds. Another appealing feature of this mechanism if the fast reheating, or at least the fast particle production associated with tachyonic preheating. This ensures the quick production of a radiation bath (provided thermalization is sufficiently fast) that can sustain the warm inflaton's slow-roll evolution at a relatively high temperature, with strong dissipative effects.

For warm inflation we chose the realization within the ``Warm Little Inflaton'' \cite{Bastero-Gil:2016qru} setup since this is the best studied model in terms of not only the underlying microphysics but also the background and perturbation dynamics. Other realizations of both cold and warm inflation (e.g.~``minimal warm inflation'' with an axion-like field \cite{Berghaus:2019whh}) may of course be possible, but their study is left for future work.

Our working example has an approximate 30-20-10 splitting of the last 60 e-folds of inflation in terms of the cold-warm-thermal stages. Other combinations are of course possible, but we note that a shorter period of cold inflation should be harder to reconcile with observations, while a longer thermal inflation stage is challenging from the dynamical perspective, since inflation is in this case sustained by an exponentially dropping temperature.

This splitting has an important impact on structure formation, since the small-scale curvature perturbations that become super-horizon during the warm inflation stage have an enhanced amplitude due to thermal dissipation, compared to the amplitude of the large-scale perturbations crossing the horizon during the first cold inflation period. In our working example the power spectrum is enhanced by 7 orders of magnitude, despite the lower Hubble scale during warm inflation. This means that light PBH formation is natural within our setup. 

Here we should pause to note that our proposal is not, of course, the first to consider multiple epochs of inflation, and other multi-stage inflation scenarios have in fact been proposed as a means to produce PBHs \cite{Kawasaki:1997ju, Kawasaki:1998vx, Kawasaki:2006zv, Kawaguchi:2007fz, Frampton:2010sw, Kawasaki:2012kn, Kawasaki:2016ijp, Kawasaki:2016pql, Inomata:2016rbd, Inomata:2017okj, Inomata:2017uaw, Tada:2019amh}. The novelty of our setup lies in the distinct dynamics of the fields driving inflation in the different epochs, with the advantages of (1) requiring only a single scalar direction with $m\lesssim H$ and (2) the intermediate warm inflation stage naturally amplifying the small-scale curvature power spectrum.

The PBH mass function that results from our trilogy setup is, in fact, quite peculiar. On the one hand, the slightly blue-tilted curvature power spectrum that characterizes warm inflation in the strong dissipation regime naturally favours the formation of sub-solar mass PBHs; on the other hand, the lightest PBHs form before thermal inflation, and as a consequence their abundance is greatly diluted. This means that light but cosmologically stable PBHs are a natural prediction of our scenario, potentially explaining all the dark matter in the universe. Thermal inflation also leaves a broad gap in the PBH mass spectrum, as first observed in \cite{Green:1997sz}. We note that in the original WLI setup the warm inflaton is also stable at late times \cite{Rosa:2018iff, Levy:2020zfo}, so that its remnant may also account for a fraction of dark matter in this particular realization of our scenario. 

The main prediction of our setup is thus a gapped PBH mass function with an underlying nearly scale-invariant, blue-tilted curvature power spectrum, while the large-scale curvature power spectrum may be red-tilted in agreement with CMB data. The PBH abundance in the sub-solar mass regime is also expected to be significant in general, so that our scenario could potentially be probed with future observations, particularly microlensing surveys (or other potential probes of the asteroid-mass window). 

The ``mini-PBHs'' predicted by our model are likely harder to probe, depending on how much their abundance is diluted by thermal inflation, but there are certainly parametric regimes where the effects of their evaporation on the galactic and extra-galactic gamma-ray backgrounds can be probed, and which we plan to study in detail in future work. Another potential way of probing these mini-PBHs is through the background of relic gravitons resulting from their mergers in the early Universe, along the lines proposed in \cite{Hooper:2020evu}.

\begin{acknowledgments}
This work was supported by FCT - Funda\c{c}ão para a Ci\^encia e Tecnologia, I.P. through the projects CERN/FIS-PAR/0027/2021,
UIDB/04564/2020 and UIDP/04564/2020,
with DOI identifiers 10.54499/CERN/FIS-
PAR/0027/2021, 10.54499/UIDB/04564/2020 and
10.54499/UIDP/04564/2020, respectively. P.B.F. is supported by the FCT fellowship SFRH/BD/151475/2021 with DOI identifier 10.54499/SFRH/BD/151475/2021. 
\end{acknowledgments}


\appendix

\section{Slow-roll conditions}

\label{appendix1}

In this appendix we review the conditions for the slow-roll of the cold and warm inflaton fields in the relevant inflationary epochs.

We start by discussing the dynamics of the dominant scalar field $\phi$ driving inflation, in the most general case including a friction term $\Upsilon\dot\phi$ in the equation of motion. In this case, in the slow-roll regime we have:
\begin{equation}
3H(1+Q)\dot\phi\simeq -V'(\phi)~, \qquad H^2\simeq {V(\phi)\over 3M_P^2}~,
\end{equation}
with $Q=\Upsilon/3H$. Then, differentiating the Friedmann equation and using the field equation we obtain:
\begin{equation}
2H\dot{H}\simeq {V'\dot\phi\over 3M_P^2}\simeq -{V'^2\over 9HM_P^2(1+Q)}
\end{equation}
so that
\begin{equation}
\epsilon_H=-{\dot{H}\over H^2}\simeq {V'^2\over 18H^4M_P^2(1+Q)}={\epsilon_\phi\over 1+Q}
\end{equation}
where $\epsilon_\phi={M_P^2\over 2}(V'/V)^2$ and in the last step we have used the Friedmann equation. This means that slow-roll inflation, $\epsilon_H<1$, implies $\epsilon_\phi<1+Q$ for warm inflation, which in the absence of dissipation reduces to the more conventional (cold inflation) condition, $\epsilon_\phi<1$. Further requiring a slow change of the $\epsilon_H$ parameter, i.e.~$\dot{\epsilon}_H/\epsilon_H<H$, it is easy to show that the condition $|\eta_\phi|<1+Q$ must also be satisfied, with $\eta_\phi=M_P^2 V''/V$.

Let us consider now the case where $\phi$ is not the dominant scalar field driving inflation. In this case, the relevant slow-roll equations (in the absence of dissipation since this is the case we are mostly interested in) are:
\begin{equation} \label{SR_subdominant}
3H\dot\phi\simeq -V'(\phi)~,\qquad H^2\simeq {\rho_{inf}\over 3 M_P^2}
\end{equation}
where $\rho_{inf}$ denotes the energy density of the dominant field. A slow-roll trajectory for the sub-dominant field then implies, using Eqs.~(\ref{SR_subdominant}):
\begin{equation}
{{1\over2}\dot\phi^2\over V}\simeq \epsilon_\phi {V\over 3\rho_{inf}}={\epsilon_\phi\over 3} \Omega_\phi <1
\end{equation}
or equivalently $\epsilon_\phi<3\Omega_\phi^{-1}$, such that the slow-roll condition is relaxed for a sub-dominant scalar field, $\Omega_\phi<1$. An analogous condition for $|\eta_\phi|$ results from requiring $|\ddot\phi|< 3H|\dot\phi|$.

\section{Spectral index for warm inflation}
\label{sisection}
The spectral index can be computed from the curvature perturbation power spectrum as
\begin{align}
n_s-1 = \frac{d\log \Delta_{\mathcal{R}}^2}{d\log k} \simeq \frac{d\log \Delta_{\mathcal{R}}^2}{dN_e}.
\end{align}
For a general warm inflation model, the curvature perturbation power spectrum can be written as \cite{Ramos:2013nsa}:
\begin{align}
&\Delta_{\mathcal{R}}^2 =\frac{V_*(1+Q_*)^2}{24\pi^2M_P^4\epsilon_\phi}\Big(1+2n_*+\frac{2\sqrt{3}\pi Q_*}{\sqrt{3+4\pi Q_*}}\frac{T_*}{H_*}\Big)G(Q_*)=\frac{V_*(1+Q_*)^2}{24\pi^2M_P^4\epsilon_\phi}F(Q_*)~,\\
&G(Q_*) \simeq 1+\alpha_wQ_*^{\beta_w}+\alpha_\nu Q_*^{\beta_\nu}.
\end{align}
where starred quantities are evaluated when the corresponding scales cross the horizon during inflation. For a linear dissipation coefficient, $\Upsilon=C_T T$ we have $Q=\Upsilon/3H=(C_T/3)(T/H)$. In the slow-roll regime, we have $3H(1+Q)\dot\phi\simeq -V'$ and $4H\rho_R\simeq \Upsilon\dot\phi^2$, and we may write $\rho_R=C_RT^4$ for a near-equilibrium radiation bath. Combining these we obtain the relation:
\begin{equation} \label{Q_phi}
Q^3(1+Q)^2={C_T^4\over 18C_R}\epsilon_\phi {M_P^4\over V}~.    
\end{equation}
Differentiating both sides of this equation with respect to the number of e-folds we then obtain the slow-roll equation for the dissipative ratio $Q$:
\begin{equation}\label{SR_Q}
{Q'\over Q}={6\epsilon_\phi-2\eta_\phi\over 3+5Q}~.  
\end{equation}
We may then use this to compute the scalar spectral index, yielding:
\begin{align}
n_s-1 \simeq \frac{2\eta_\phi-6\epsilon_\phi}{1+Q_*}\Big(1-\frac{2Q_*}{3+5Q_*}-\frac{Q_*(1+Q_*)}{3+5Q_*}\frac{d\log F(Q_*)}{dQ_*}\Big).
\end{align}
In the strong dissipation regime we have $F(Q_*)\simeq \text{const.}\times Q_*^{3/2+\beta_\nu}$, assuming $\beta_\nu>\beta_w$. Then, for a quartic potential:
\begin{align}
n_s-1 \simeq -\frac{24M_P^2}{\phi_*^2}Q_*^{-1}\Big(\frac{3}{10}-\frac{\beta_\nu}{5}\Big),
\end{align}
where we considered a quartic potential for the warm inflaton. We can integrate Eq.~(\ref{SR_Q}) using (\ref{Q_phi}) to give the total duration of the warm inflation stage:
\begin{align}
N_e \simeq \frac{15}{2}\frac{\phi_*^2}{24M_P^2}Q_*,
\end{align}
such that 
\begin{align}
n_s-1\simeq -\frac{15}{2N_e}\Big(\frac{3}{10}-\frac{\beta_\nu}{5}\Big).
\end{align}
For $\beta_\nu=2.315$ as obtained numerically for a quartic potential \cite{Bastero-Gil:2016qru} and setting $N_e=22$ as in our working example, this yields $n_s=1.056$.


\end{document}